\documentstyle[12pt]{article}
\begin{document}
\def\be{\begin{eqnarray}}
\def\en{\end{eqnarray}}
\def\t{\times}
\def\non{\nonumber}
\def\n{\nonumber \\}
\def\la{\langle}
\def\ra{\rangle}
\def\ep{\varepsilon}
\def\u{{\mu}_{\rm fact}}
\def\ums{{\mu}_{\rm MS}}
\def\lsim {{\ \lower-1.2pt
\vbox{\hbox{\rlap{$<$}\lower5pt\vbox{\hbox{$\sim$} }}}\ } }
\def\gsim  {{\ \lower-1.2pt\vbox{\hbox{\rlap{$>$}\lower5pt\vbox{\hbox{$\sim$}
}}}\ } }
\def\pr{{\sl Phys. Rev.}~}
\def\prl{{\sl Phys. Rev. Lett.}~}
\def\pl{{\sl Phys. Lett.}~}
\def\np{{\sl Nucl. Phys.}~}
\def\zp{{\sl Z. Phys.}~}
\def\jp{{\sl J. Phys.}~}

\font\el=cmbx10 scaled \magstep2 {\obeylines
\hfill IP-ASTP-3-97
\hfill April, 1997}
\vskip 1.5cm
{\obeylines
\parskip 27pt
\baselineskip 27pt
\centerline{\large \bf Double parton scattering of hadron-hadron interaction}
\centerline{\large \bf and its gluonic contribution}}
\vskip 3.5cm
\centerline{Hung Hsiang Liu \footnote{E-mail: hliu@phys.sinica.edu.tw}}

\vskip 25pt {\obeylines
\parskip 15pt
\baselineskip 15pt
\centerline{Institute of Physics, Academia Sinica}
\centerline{Nankang, Taipei, Taiwan 115, R.O.C.} }
\vskip 2cm

\centerline{\bf Abstract}
\bigskip
\bigskip  {\small We propose a formalism for calculating the event cross-section
of parton interactions. In this formalism, we use the light-front bound-state
wave function to expand scattering initial state. This leads to an expression of
the  cross-section in terms of square of wave functions which can be used to define the parton distribution functions. The probability interpretation of the partonic
interactions among hadron-hadron scattering is  naturally achieved.  We
apply this formalism to calculate the 4-particle final state cross-section for  single parton, double parton and gluon splitting interactions. We compare
the behaviors of these cross sections and propose a method for experiments to
differentiate these interactions.}
\newpage   
\noindent{\it\bf I. Introduction}  
\vskip 0.3cm
    Phenomenological calculation of  hadron-hadron cross section  uses
the  probability of  finding  one parton in each scattering hadron multiplied
by the  parton-level cross section $d{\hat\sigma}$:
\be d \sigma (A+B \rightarrow {\rm n}~{\rm jets}+{\rm anything})&=&\sum_{a,b}
\int  dx_a dx_b  F_{a/A}(x_a) F_{b/B}(x_b)\non \\ &\t& d {\hat \sigma} (a+b
\rightarrow {\rm n}~{\rm jets}).
\label{eq: ddab}
\en In the calculation, one uses  universal parton distribution functions for
$F$  and uses perturbative QCD to calculate the n-jet single parton cross
section
${\hat \sigma}_{2\rightarrow n}$.  Since 1982 various experiments studied two-, three- and four-jet events from  hadron-hadron scattering \cite{ua1}.  The two- and
three-jet cross sections agree well with theoretical calculations \cite{3j}. However, the
experiments on four-jet events show some inconsistence to the calculations \cite{4j}.
The UA2  \cite{ua2} experiment    showed a good agreement to the calculations
based on single parton interaction. On the other hand the $p{\bar p}$ collision
experiments by  AFS Collaboration at CERN ISR \cite{afs} and the CDF experiments \cite{cdf} 
showed a small violation to the calculations. Consequently CDF and AFS
collaborators  suggested that a small double parton contribution to the QCD $(2
\rightarrow 4 jets)$ cross section may provide a better fit to the experimental
data.  

The double parton interaction happens when two partons in one hadron collide
with  two partons in another hadron [Fig.~1].  Since this process
involves more than one parton in a hadron at one time, it will give more informations on hadronic structure than the single parton scattering
can give. Experimentally one can differentiate the double parton events
from single parton events by the    final jet configurations. First,
since the transverse momenta of partons in a hadron are small,  the
transverse momenta of the two jets produced in each of the parton-parton
scattering should be well balanced. Therefore in a double parton scattering,
one can find two pairs of jets with transverse momenta, say $({\bf p}_1,{\bf
p}_2)$ and $({\bf p}_3,{\bf p}_4)$, such that $|{\bf p}_1+{\bf p}_2|$, $|{\bf p}_3+{\bf p}_4|<<$1 GeV. In
the contrast, in an ordinary four-jet event one will have ${\bf p}_1+{\bf
p}_2+{\bf p}_3+{\bf p}_4\approx$0, but for any pair of transverse momenta
$({\bf p}_i,{\bf p}_j)$, one will normally find  $|{\bf p}_i+{\bf p}_j|>>$1 GeV.
Secondly, for double parton scattering, if we assume that there should not be
strong  correlation between the two partons in the individual colliding hadron,
the distribution of the double parton cross section in the  azimuthal angle 
between the jet pairs should be flat. The CDF collaboration  defines two
variables which are sensitive to these signals.
They define variable $S$  to exploit the tendency of pairwise
balance,
\be
 S(i+j,k+l)= min\sqrt {{1\over 2}[({|{\bf p}_i+{\bf p}_j|\over \sqrt {|{\bf
p}_i|^2+|{\bf p}_j|^2} })^2+({|{\bf p}_k+{\bf p}_l|
\over \sqrt {|{\bf p}_k|^2+|{\bf p}_l|^2} })^2]}, 
\en where $S$ is minimized over the three possible jet pairings 
$(12,34)$, $(13,24)$ and $(14,23)$ and they define the variable $\Delta S$  
to be the azimuthal angle between the two jet pairs which minimize $S$. The
double parton events should favor smaller values of $S$ and the single
parton events should favor larger ones. Also, the double parton events should be uniformly
distributed over  $\Delta S$ in the interval $0$ to $\pi$. 

In this paper we  discuss the theory of the 4-particle final state events from hadron-hadron  scattering. Most of these events are in the form of 4-jet production. There is a small fraction of these events which is in the form of di-lepton pair production. It is believed that the hadron fragmentation of the jet final state, which is a soft  process, does not change the kinematics of the final particles. Therefore the di-lepton pair production  should have a
similar kinematics to the jet production. For a theoretical  understanding  of 
double parton scattering, we will use the di-lepton process as  examples to the formalism. The same formalism can be applied to calculate the jet cross 
section. We are interested in the experiments with observables $Q_1,
Q_2$ of the TeV order, where $Q_1$ and $Q_2$ are the invariant momenta of the final particle pairs. The largest source of these events is the lowest order single parton scattering. Double parton scattering may contribute these events. There is also a possible type of four-particle events that a gluon splits before collision. In this process, one incoming gluon splits into two partons, while the other incoming gluon also
splits, then the two pairs of partons collide. One can interpret this process as
an alternative form of double parton scattering. Imagining that the parton pairs
are produced in time $\Delta t$  before the collision, where  $\Delta t$ is long compared to the
parton collision time $1/Q$  but short compared to the typical time for  partons within a hadron to interact, then the transverse momenta of the partons  produced in the gluon splitting process will be in the
rang of 0$<<{\bf k}<<Q_1,Q_2$. We will examine the effect of these large transverse momentum partons on  double parton scattering. 
The paper is organized in the following fashion. In the next section, we
 introduce the light-front wave functions which are useful for the formalism. In section III, we derive the cross section for the single parton scattering. In
section IV, we use the formalism to calculate the double parton cross section. 
 In section V, we discuss the gluon
splitting scattering. Finally, in the discussion section we will compare the
four-particle final state cross sections of single parton scattering, double parton scattering and from gluon splitting scattering. We propose a method for experiments to differentiate these interactions.
\newpage
\noindent{\it\bf II. Light-front wave functions}
\vskip 0.3cm
   In the light-front field
theory, one uses a Hamiltonian approach to derive the equation for a bound-state
wave function. For this purpose, one uses gauge field theory quantized on
light-front planes. That is, one attempts to solve the eigenvalue equation
$P^-_{\rm op}|\Psi\rangle=E|\Psi\rangle$, where $P^-_{\rm op}$ is the generator
of translations from one light-front plane $x^+={\rm const}$ to the next. 
Consider a state  $|P^{+},{\bf P}\rangle$ which denotes  a hadron with light-front momentum 
$(P^+,({\bf P}^2+M^2)/(2P^+),{\bf P})$,
normalized by
$\langle P^+,{\bf P}|P^{'+},{\bf P'}\rangle=(2\pi)^3
2P^+\delta(P^+-P^{'+})\delta^2({\bf P}-{\bf P'}).
$\\
We can expand the state vector in the Fock space
\be |P^+,{\bf P}\rangle&=&\sum_m \sum_{s_1,s_m}(\prod_{i=1}^m{\int dp^+_id{\bf
p}_i\over (2\pi)^3 })b^+(p^+_1,{\bf p}_1;s_1)..d^+(p^+_m,{\bf p}_m;s_m)|0\rangle
\n &\t& \psi_{m}(p_1,..p_m;s_1,...s_m) 2P^+(2\pi)^3\delta(P^+-\sum_{i=1}^m p_i^+)
 \delta^2({\bf P}-\sum_{i=1}^m {\bf p}_i).\label{eq: pmpp}
\en
The advantage of this framework that one hopes will make the problem tractable
is that partons (at least those with zero momentum) cannot appear spontaneously
from the vacuum: 
The  internal state of the hadron is given in terms of a set of wave functions
$\psi_n$ defined by
\be &&\langle 0|b(p_1^+,{\bf p}_1;s_1)d(p_2^+,{\bf p}_2;s_2)...d(p_m^+,{\bf
p}_m;s_m)|P^+,{\bf P}\rangle\n &&=(2\pi)^32P^+ \delta(P^+-\sum_{i=1}^m
p_i^+)\delta^2({\bf P}-\sum_{i=1}^m {\bf p}_i^+) \psi_n(p_1^+,{\bf
p}_1;p_2^+,{\bf p}_2...p_m^+,{\bf p}_m;s_1,..s_m),
\label{eq: bpps}\en where $b$'s and $d$'s are the destruction operators for
quarks and antiquarks. The indices $s_i$  in the wave functions are the internal 
symmetry variables which denote helicity $\lambda_i$,  isospin $t_i$,  and
color $c_i$.
From the normalization equation we can show
\be 1&=&\sum_m \sum_{s_1.. .s_m}(\prod_{i=1}^m {\int dp_i^+ d^2{\bf p}_i \over
(2\pi)^{3} 2p_i^+})(2\pi)^32P^+\delta(P^+-\sum_{i=1}^m p_i^+)\delta^2({\bf
P}-\sum_{i=1}^m{\bf p}_i)  \n   &\times&|\psi_m(p^+_1,{\bf p}_1,...p^+_m,{\bf
p}_m;s_1,...s_m)|^2.\label{eq: mssm}
\en

The normal light-front wave function is good for describing particles moving in
the $+z$ direction. In order to describe  the bound-state particles moving  in the
$-z$ direction, we have  to do the followings: 
We expand the state vector for hadron moving in the $-z$ direction in the    Fock space by
\be |P^-,{\bf P}\rangle&=&\sum_n \sum_{s_1,s_n}(\prod_{i=1}^n{\int dp^-_id{\bf
p}_i\over (2\pi)^3 })b^+(p^-_1,{\bf p}_1;s_1)..d^+(p^-_n,{\bf p}_n;s_n)|0\rangle
\n &\t& \psi_{n}(p_1,..p_n;s_1,...s_n) 2P^-(2\pi)^3\delta(P^--\sum_{i=1}^n p_i^-)\delta^2({\bf P}-\sum_{i=1}^n {\bf p}_i).\label{eq: pnpp}
\en
Define the  spinors  $\mu(p,s)$ and  $\nu(p,s)$ as the  spin components of 
quarks and anti-quarks moving in the $-z$ direction:
\be
\mu(p,1/2)={1\over {\sqrt {\sqrt 2}p^-}}\pmatrix{p^1-ip^2\cr {\sqrt 2}p^-\cr
0\cr m}, \quad\quad \mu(p,-1/2)={1\over {\sqrt {\sqrt 2}p^-}}\pmatrix{m\cr
0\cr {\sqrt 2}p^-\cr -p^1-ip^2},\non 
\en
\be
\nu(p,1/2)={1\over {\sqrt {\sqrt 2}p^-}}\pmatrix{-m\cr 0\cr {\sqrt 2}p^-\cr
-p^1-ip^2}, \quad\quad 
\nu(p,-1/2)={1\over {\sqrt {\sqrt 2}p^-}}\pmatrix{p^1-ip^2\cr {\sqrt 2}p^-\cr
0\cr -m}.\label{eq: pppp}
\en 
The normalization of these vectors is:
\be
\sum_s \mu_{\alpha}(p^-,{\bf p},s)\overline{\mu}_\alpha(p^-,{\bf p},s)=
-\sum_s \nu_{\alpha}(p^-,{\bf p},s)\overline{\nu}_a(p^-,{\bf p},s)=p\!\!\!\slash+m.\label{eq: spps} 
\en
It can be shown that   these spinors are the
eigen-states of a helicity operator referring to a Lorentz frame moving in the
$+z$ direction near  the speed of light, $i.e.$, $v_z=-{\rm tanh}(\omega),
\omega\rightarrow -\infty$ \cite{soper}.  The helicity operator is
\be h_{\omega}(p)\rightarrow h_{-\infty}(p)={1\over {\sqrt
2}p^-}\pmatrix{{-p^-\over {\sqrt 2}}&p^1-ip^2&0&0\cr 0&{p^-\over {\sqrt
2}}&0&0\cr 0&0&{-p^-\over {\sqrt 2}}&0\cr 0&0&p^1+ip^2&p^-\over {\sqrt 2}}.
\label{eq: hphp}\en 
One can check that  the  spinors listed in Eq. (\ref{eq: pppp})  are  eigen states of  $h_{-\infty}(p)$:
\be h_{-\infty}(p)\mu(p,\pm {1\over 2})=\pm {1\over 2}\mu(p,\pm {1\over
2}),\n h_{-\infty}(p)\nu(p,\pm {1\over 2})=\mp {1\over 2}\mu(p,\pm {1\over
2}).
\label{eq: hhpp}\en
We therefore clarify that  the physical meaning of  the spin index $s$ is the
helicity of the quark as measured in a Lorentz frame moving near the speed of
light in the  $+z$ direction. Similarly, one can define the photon polarization
vectors $\epsilon^\mu(\lambda)$ as
\be
\epsilon^\mu(\pm 1)={1\over \sqrt 2}({q^1\pm iq^2\over q^-}, 1, \pm i, 0).
 \label{eq: qqqi}\en The physical meaning of the index $\lambda$ is the 
helicity of the photon as measured in a frame moving the $+z$ direction. 
\newpage
\noindent {\bf III.~The cross section for single parton scattering}
\vskip .3cm
   In this section we derive the cross section formula for single parton
scattering. For simplicity, we use the di-lepton production from  unpolarized
hadron-hadron scattering as an example.  We are interested in the experiments with observables $Q_1$, $ Q_2$ of the TeV order. The transverse momenta of partons in a hadron are therefore negligible. The differential  cross section for this process as shown in Fig.~2 is
\be {d\sigma_{sp} \over{d^4Q_1 d^4Q_2}} &=&{1\over 2s}{1\over 2}\sum_{S_{A}}{1\over
2}\sum_{S_{B}}  \prod_{i=1}^4 {d^3l_i \over 2l_i^0(2\pi)^3}
|{\rm M|^2_{sp}}\delta^4 (l_1+l_2-Q_1)\delta^4 (l_3+l_4-Q_1)\n   &\times&
(2\pi)^4\delta^4(P_A+P_B-\sum_{i=2}^m a_i-\sum_{i=2}^n b_i-Q_1-Q_2),
\label{eq: dpap}
\en where $Q_1$ and $Q_2$ are the invariant 4-momenta of the lepton pairs ($l_1$, $l_2$) and ($l_3$, $l_4$) and $|{\rm M}|$ is the invariant hadronic scattering amplitude which is
calculated by using its relation to  the scattering matrix
\be  _{\rm out}\langle {\rm
anything},l_1,l_2,l_3,l_4|S|A,B\rangle_{\rm in}=(2\pi)^4\delta^4(p_i-p_f){\rm M}.
\label{eq: outa}
\en 
The state $|A,B\rangle_{\rm in}$ denotes the initial state for hadron $A$ moving in the $+z$ direction and hadron $B$ moving in the $-z$ direction. We assume that the  state  $|A,B\rangle_{\rm in}$ can be factorized into $|A\rangle_{\rm in} |B\rangle_{\rm in}$. Then we expand the hadron state $|A\rangle_{\rm in}$
 by Eq. (\ref{eq: pmpp}) and expand the hadron state $|B\rangle_{\rm in}$ by Eq. (\ref{eq: pnpp})
After applying these expansions we can express the scattering amplitude square as
\be |{\rm M}|_{\rm sp}^2 &=& \sum_{m,n}\sum_{s_{a2}..s_{an}}\prod_{i=2}^m
\int {da^+_i d^2{\bf a}_i\over (2\pi)^3 2a_i^+}     
\sum_{s_{b2}..s_{bn}}\prod_{i=2}^n\int {db^-_i d^2{\bf b}_i\over (2\pi)^3
2b_i^-}|{\rm m}|^2\non \\   &\t&{1\over  x^2_{a1}}  
\psi^2_{am}(x_{a1},-\sum_{i=2}^m{\bf a}_i,s_{a1};x_{a2},{\bf
a}_2,s_{a2}...;x_{am},{\bf a}_m,s_{am})\n  &\t&{1 \over  x^2_{b1}}\psi^2_{bn}(x_{b1},{\bf
Q}_1+{\bf Q}_2+\sum_{i=2}^m{\bf a}_i,s_{b1}; x_{bn},{\bf b}_2,s_{b2}...;b_n,s_{bn})\n  &\t&
\int  d^2 {\bf p}_a \delta^2({\bf p}_a-\sum_{i=2}^m {\bf a}_i).
\label{eq: mxxd}
\en 
The wave
function $\psi_{am}$($\psi_{bn}$) is a function of the hadron spin
$S_{A}$($S_{B}$), but we do not explicitly display this dependence.
We inserted an identity $1=\int  d^2 {\bf p}_a \delta^2({\bf
p}_a-\sum_{i=2}^m {\bf a}_i)$ in Eq. (\ref{eq: mxxd}) for future purpose. This
insertion serves to define   ${\bf p}_a$ as the total transverse momentum of
the non-interacting partons  numbered from $2$ to $m$. The parton momentum fractions are defined by $x_{ai}\equiv a_i^+/P_A^+$ and $x_{bi}\equiv b_i^-/P_B^-$. The momentum fractions of the interacting partons $a_1$ and $b_1$ are fixed by the overall delta function. They are 
\be x_{a1}&=&{P_A^+-\sum_{i=2}^m a_i^+\over P_A^+}={Q_1^++Q_2^+\over
P_A^+}+O({M_B^2\over s},{{\bf b}_i^2\over s}),\n x_{b1}&=&{P_B^--\sum_{i=2}^n
b_i^-\over P_B^-}={Q_1^-+Q_2^-\over P_B^-}+O({M_A^2\over s},{{\bf a}_i^2\over
s}).\label{eq: xxbp}
\en  
In this equation $|{\rm m}|_{\rm sp}^2$ is the invariant partonic scattering amplitude square. Its integration
over the final particle phase space denoted by $|T|_{\rm sp}^2$  will be calculated in
the appendix.  
In order to simplify the formula, we  collectively express the sets of the
non-interacting parton  momenta and spins by:
\be V_a&=&{\large \{}(a_2^+,{\bf a}_2;...a_m^+,{\bf a}_m){\large \vert}\sum^m_{i=2}
a_i^+=P_A^+-Q^+_1-Q^+_2,\sum_{i=2}^m{\bf a}_i= {\bf p}_a {\large \}},\n  
V_b&=&{\large \{}(b_2^-,{\bf
b}_2;...b_n^+,{\bf b}_n){\large \vert}\sum^n_{i=2} b_i^-=P_B^+-Q^-_1-Q^-_2,
\sum_{i=2}^n{\bf b}_i= -{\bf p}_a-{\bf Q}_1-{\bf Q}_2{\large \}},\non \\ 
S_a&=&\{s_{a2},...s_{am}{\large \vert}\sum^m_{i=1}s_{ai}=S_A\}, \non \\
 S_b&=&\{s_{b2},...s_{bn}{\large \vert}\sum^n_{i=1}s_{bi}=S_B\},\label{eq: vvss}
\en
and we denote the phase space of the non-interacting final partons as
\be
\int[dV_a]&=&\prod_{i=2}^m\int{ da_i^+d^2{\bf a}_i \over (2\pi)^3 2a_i^+}
2P_A^+(2\pi)^3  \delta(P_A^+-\sum_{i=2}^m a^+_i-Q^+_1-Q^+_2)\delta^2 ({\bf
p}_a-\sum_{i=2}^m {\bf a}_i),\non \\
\int[dV_b]&=&\prod_{i=2}^n\int{db_i^-d^2{\bf b}_i \over (2\pi)^3 2b_i^-}
2P_B^-(2\pi)^3  \delta(P_B^--\sum_{i=2}^n b^-_i-Q^-_1-Q^-_2)\non \\
&\t& \delta^2({\bf
p}_a+{\bf Q}_1+{\bf Q}_2+\sum_{i=2}^n {\bf b}_i).\label{eq: ddpa}
\en  
With these notations we can rewrite Eq. (\ref{eq: dpap}) as
\be  {d\sigma\over {d^4Q_1 d^4Q_2}}&=&{d\sigma\over {d^2{\bf Q}_Td^2\Delta{\bf
Q}dQ^+_1 dQ^-_1dQ^+_2  dQ^-_2}}\n 
 &=&{1\over 4s^2(2\pi)^2}{1\over x^2_{a1}x^2_{b1}} {1\over
2}\sum_{S_A}\sum_{S_{a}}{1\over 2}\sum_{S_B}\sum_{S_{b}}\int[dV_a]\int
[dV_b]\non \\ &\t&
\sum_{m,n}\sum_{s_{a1},s_{b1}} |T|^2 \int d^2{\bf p}_a \int d^2{\bf a}_1\delta^2({\bf
p}_a+{\bf a}_1)\non \\ &\t&   
\psi^2_{am}(x_{a1},{\bf a}_1,s_{a1};{\bf p}_a;V_a,S_a)\non \\
&\t&\psi^2_{bn}(x_{a1},{\bf Q}_T-{\bf a}_1,s_{b1};-{\bf Q}_T+{\bf p}_a;V_b,S_b) 
,\label{eq: dsma}
\en 
 in which we define ${\bf Q}_T={\bf Q}_1+{\bf Q}_2$, $\Delta {\bf Q}=({\bf Q}_1-{\bf Q}_2)/2$. For the following discussion we inserted an identity    
$1=\int d^2{\bf a}_1 \delta^2 ({\bf a}_1+{\bf p}_a)$ in Eq. (\ref{eq: dsma}) The wave
function $\psi_{am}$ (or  $\psi_{bn}$) in Eq. (\ref{eq: dsma}) is the same wave function in Eq. (\ref{eq: mxxd}) expressed by
the new variables ${\bf a_1},{\bf p}_a, V_a$(or  ${\bf b_1},{\bf p}_b, V_b$). We can
use these momenta to define the following variables:
\be &&\Delta{\bf p}_a= x_a {\bf a}_{1}- x_{a1}{\bf p}_a  \non \\  &&{\bf A}={\bf
p}_a+{\bf a}_{1},
\label{eq: papa}
\en  in which   $x_a$ is the total  momentum fraction  of the non-interacting
partons and ${\bf A}$ is the total transverse momentum of hadron A. 
By using coordinate ${\bf r}_{a1}$ of parton $a_1$  and the center  mass coordinate ${\bf r}_a$ of the remaining partons, we can define the following variables  which  conjugate to 
momenta  $\Delta{\bf p}_a$ and ${\bf A}$:
\be &&\Delta{\bf r}_a={\bf r}_{a1}-{\bf r}_{a},\n  
 &&{\bf R}_a=x_{a1}{ {\bf r}}_{a1}+ x_{a}{\bf r}_a
,\label{eq: rrax}\en   
 here   $\Delta{\bf r}_a$ is the relative transverse   distance between parton
$a_1$ and the center mass of the non-interacting partons and ${\bf R}_a$ is the
center mass coordinate of hadron A.   Similarly, we can   define 
the   momentum variables $\Delta{\bf p}_b$ and ${\bf B}$ and  their conjugate
coordinates $\Delta{\bf r}_b$  and ${\bf R}_b$  for hadron B.     To
express the cross section Eq. (\ref{eq: dsma})  in terms of these  coordinate variables,
we  first express the wave functions in terms of $\Delta{\bf p}_a$ and ${\bf
A}$  then  Fourier transform these wave functions to the coordinate space 
by using, for example,
$\non \\ \psi(\Delta{\bf p}_a,{\bf A};V_A)={1/ 2\pi}\int  d^2\Delta{\bf r}_a  {\rm
exp}(i\Delta{\bf p}_a\cdot\Delta{\bf r}_a){\tilde \psi}(\Delta{\bf r}_a,{\bf A};V_A)$.
 
The integration volume $\int d^2{\bf p}_a \int d^2{\bf a}_1$ in Eq. (\ref{eq: dsma}) can be
replaced by $\int d^2{\bf p}_{a21}\int d^2{\bf A}$. It is easy to see from the overall delta function that 
the total transverse momentum ${\bf Q}_T$ is of the order of the parton transverse momentum which is of a few MeV. Since we are interested in the experiments with observables $Q_i^+$, $Q_i^-$ of the TeV order, we can neglect the ${\bf p}_{a21}$,  ${\bf A}$ and ${\bf Q}_T$ dependence in $|T|^2$ and we can integrate Eq. (\ref{eq: dsma}) over these transverse momenta.
The integration of ${\bf Q}_T$ gives us a delta function which enables us to express the differential cross section in terms of   the wave function squares:
\be  {d\sigma\over {d^2\Delta {\bf Q}dQ^+_1 dQ^-_1
dQ^+_2dQ^-_2}}&=&{1\over  4s^2(2\pi)^2}{1\over x^2_{a1}} {1\over x^2_{b1}}
\sum_{m,n}\sum_{s_{a1} s_{b1} } |T|^2 \non\\  &\t&{1\over 2}\sum_{S_A}\sum_{S_{a}}\int
[dV_a] \int d^2\Delta{\bf r}_{a{\bar a}} \non |{\tilde \psi}_{am}(x_{a1},s_{a1};\Delta{\bf
r}_{a{\bar a}};V_a,S_a)|^2\n &\t& {1\over 2}\sum_{S_B}\sum_{S_{b}}\int [dV_b]
\int d^2\Delta{\bf r}_{b{\bar b}} |{\tilde \psi}_{bn}(x_{b1},s_{b1};\Delta{\bf r}_{b{\bar
b}};V_b,S_b)|^2. 
\label{eq: dssb}
\en  
We can define the following function $f_a$ (and similarly $f_b$):
\be f_a(x_{a1} ,s_{a1})&=&{1\over (2\pi)^3}{1\over 2x_{a1}}{1\over 2} 
\sum_{m}\sum_{S_{A}}\sum_{S_{a}}\int [dV_a]\int d^2\Delta{\bf r}_{a{\bar a}} \non
\\ &\t&|{\tilde \psi}_{am}(x_{a1},s_{a1},\Delta{\bf r}_{a{\bar a}}; V_a,S_a)|^2.\non \\
\label{eq: famx}\en 
From the normalization equation Eq. (\ref{eq: mssm}) , one sees that 
\be \sum_{s_{a1}}\int dx_{a1} f_a(x_{a1} ,s_{a1})=1. 
\n
\en  
The function $f_a$ can then be interpreted as the distribution function for
parton $a_1$ with momentum fraction $x_{a1}$ and internal symmetric variable
$s_{a1}$. 

If we assume that for non-polarized hadron the distribution functions of different helicities are equal, $i.e.$, $f(x,1)=f(x,-1)=1/2f(x)$, we get the cross section for single parton scattering as
\be {d\sigma\over {d^2 \Delta{\bf Q }dQ^+_1 dQ^-_1 dQ^+_2dQ^-_2}}&=&
{d\sigma\over {{1\over 2}d^2\Delta{\bf Q }d{Q}^2_1 dy_1 {1\over 2}d{ Q_2}^2
dy_2}}\n 
&=&{(2\pi)^4\over 4s^2}{1\over x_{a1 } } {1\over x_{b1} }
 f_a(x_{a1}) f_b(x_{b1})|T|^2,
\label{eq: dsxa}\en  
where $y$'s are the rapidities defined by $y_i=1/2\ln(Q_1^+/Q_i^-)$ and $|T|^2$ is calculated by using the ($2\rightarrow 4$  particle) Feynman diagrams.
\vskip 0.3cm 
\noindent {\bf IV.~The cross section for double parton scattering}
\vskip 0.3cm
   In this section we  derive the cross section formula for double parton
scattering. For simplicity, we use the di-lepton production processes in unpolarized hadron-hadron scattering as an example. The process is shown as  in Fig.~1.      
By expanding the initial states as in the previous section, we can express the scattering amplitude square for double parton interaction as 
\be |{\rm M}|_{\rm dp}^2&=& \sum_{m,n}\sum_{s_{a3}..s_{an}}\prod_{i=3}^m
\int {da^+_i d^2{\bf a}_i\over (2\pi)^3 2a_i^+}     
\sum_{s_{b3}..s_{bn}}\prod_{i=3}^n\int {db^+_i d^2{\bf b}_i\over (2\pi)^3
2b_i^-}\non \\ &\times& {1\over (2\pi)^{4}4s^2}{1\over x^2_{a1}x^2_{b2}}
  \int d^2{\bf a}_1    \int d^2{\bf a}'_1 |m|_{\rm dp}^2\non \\
&\times& \psi_{am}(x_{a1},{\bf a}_1,s_{a1};x_{a2},-{\bf a}_1-\sum_{i=3}^m {\bf
a}_i,s_{a2};x_{a3},{\bf a}_3,s_{a3}...;x_{am},{\bf a}_m,s_{am})\n &\t&
\psi_{bn}(x_{b1},{\bf Q}_1-{\bf a}_1,s_{b1};x_{b2},{\bf Q}_2+{\bf a}_1+\sum_{i=3}^m
{\bf a}_i,s_{b2};x_{b3},{\bf b}_3,s_{b3}...;x_{bn},{\bf b}_n,s_{bn})\n &\t&
\psi^*_{am}(x_{a1},{\bf a}'_1,s_{a1};x_{a2},-{\bf a}'_1,-\sum_{i=3}^m {\bf
a}_i,s'_{a2};x_{a3}, {\bf a}_3,s_{a3}...;x_{am},{\bf a}_m,s_{am})\n &\t&
\psi^*_{bn}(x_{b1},{\bf Q}_1-{\bf a}'_1,s'_{b1};x_{b2},{\bf Q}_2+{\bf
a}'_1+\sum_{i=3}^m{\bf a}_i;x_{b3},{\bf b}_3,s_{b3}...;x_{n},{\bf b}_n,s_{bn}) .
\label{eq: msab}
\en
In this equation, $|{\rm m}|^2_{\rm dp}$ is the partonic scattering matrix square and the integration over its final particle phase spaces,  $|T|_{\rm dp}^2$, will be calculated in Appendix.  To calculate the jet productions from double parton scattering, one can apply Eq. (\ref{eq: msab}) and use the corresponding  $|T|^2$  for  the jet production.
The non-interacting partons are now numbered from $i=3$ to $m$ or $n$.   The momentum fractions of the interacting partons $a_1$, $a_2$, $b_1$, $b_2$  are again fixed by the overall delta function. They are
\be
x_{ai}=Q_i^+/P_A^+ , \quad  x_{bi}=Q_i^-/P_B^-, \quad  i=1,2.\label{eq: xaiq}
\en
 
In order to simplify the expression for the cross section formula, we define analogically to the previous section
the collective variables $V_a$, $V_b$, $S_a$ and $S_b$ for the non-interacting
partons. The only difference is that there are now  $m-2$ or $n-2$ non-interacting partons. Therefore we can rewrite Eq. (\ref{eq: msab}) as
\be {d\sigma\over {d^4 Q_1 d^4  Q_2}} &=&{1\over 16s^4(2\pi)^{6}}{1\over x^2_{a1}x^2_{b2}}
\non \\ &\t&{1\over
2}\sum_{S_A}\sum_{S_{a}}{1\over 2}\sum_{S_B}\sum_{S_{b}}
\int[dV_a]\int
[dV_b]\sum_{s_{a1}s_{a2}s'_{a1}s'_{a2}}\sum_{s_{b1}s_{b2}s'_{b1}s'_{b2}}|T|^2\non
\\ &\t&\int d^2{\bf p}_a \int d^2{\bf a}_1 \int d^2{\bf a}_2 \delta^2({\bf
p}_a+{\bf a}_1+{\bf a}_2)
\psi_{am}(Q_1^+,{\bf a}_1,s_{a1};Q_2^+,{\bf a}_2,s_{a2},{\bf p}_a;V_a,S_a)\non \\
&\t&\psi_{bn}(Q_1^-,{\bf Q}_1-{\bf a}_1,s_{b1};Q_2^-,{\bf Q}_2-{\bf
a}_2,s_{b2},-{\bf p}_a-{\bf Q}_1-{\bf Q}_2;V_b,S_b)\non \\ &\t& \int d^2{\bf
p}'_a \int d^2{\bf a}'_1
\int d^2{\bf a}'_2 \delta^2({\bf p}'_a+{\bf a}'_1+ {\bf
a}'_2)\psi^*_{am}(Q_1^+,{\bf a}'_1,s'_{a1};Q_2^+,{\bf a}'_2,s'_{a2},{\bf
p}_a;V_a,S_a)\non \\ &\t& \psi^*_{bn}(Q_1^-,{\bf Q}_1-{\bf a}'_1,s'_{b1};Q_2^-,{\bf
Q}_2-{\bf a}'_2,s',_{b2}-{\bf p}_a-{\bf Q}_1-{\bf Q}_2;V_b,S_b)\non \\ &\t&
\delta^2({\bf p}_a-{\bf p}'_a).
\label{eq: dsdb}
\en
We  inserted the following identities in Eq. (\ref{eq: dsdb})
for  a more symmetrical expression:
\be 1=\int d^2{\bf p}'_a \delta^2 ({\bf p}'_a-{\bf p}_a)
\int d^2{\bf a}_2 \delta^2 ({\bf a}_1+{\bf a}_2+{\bf p}_a)
\int d^2{\bf a}'_2 \delta^2 ({\bf a}'_1+{\bf a}'_2+{\bf p}'_a).\non
\en
Define the following transverse momentum variables by using parton momenta   ${\bf a}_1,{\bf a}_2$ and ${\bf p}_a$:
\be &&{\bf a}_{21}=\xi_{a1}{\bf a}_{2}-\xi_{a2}{\bf a}_{1},\n &&{\bf
p}_{a21}=x_{a}({\bf a}_{1}+{\bf a}_{2})-(x_{a1}+x_{a2}){\bf p}_{a},\n &&{\bf
A}={\bf p}_{a}+{\bf a}_{1}+{\bf a}_{2},
\label{eq: apap}\en    here we define $\xi_{a1}=x_{a1}/(x_{a1}+x_{a2})$ and 
$\xi_{a2}=x_{a2}/(x_{a1}+x_{a2})$. The variable $x_a$ is the sum of the  momentum fractions of the
partons which are not involving in the collision. Therefore we have 
$\xi_{a1}+\xi_{a2}=1$, and $x_{a1}+x_{a2}+x_{a}=1$.  The momentum ${\bf
A}$ is the total transverse momentum of hadron A.  By using the
coordinate ${\bf r}_{a1}$ of parton $a_1$, the coordinate ${\bf r}_{a2}$ of parton $a_2$ and the center
mass coordinate ${\bf r}_a$ of the remaining partons, we can define the
following variables  which  conjugate to momenta  ${\bf a}_{21},{\bf p}_{a21}$
and ${\bf A}$:
\be &&{\bf r}_{21a}={\bf r}_{a2}-{\bf r}_{a1},\n &&{\bf r}_{{\bar a}a}={\bar
{\bf r}}_{a}-{\bf r}_{a},\n &&{\bf R}_a=(x_{a1}+x_{a2}){\bar {\bf
r}}_{a}+x_a{\bf r}_a,
\label{eq: rrrx}\en     
here ${\bar {\bf r}}_a=\xi_{a1} {\bf r}_{a1}+\xi_{a2} {\bf r}_{a2}$ is the
center  mass coordinate of the two partons in hadron $A$ which are involving in the collision, 
${\bf r}_{21a}$ is the relative transverse distance between these two partons
and ${\bf r}_{{\bar a}a}$ is the relative transverse  distance between the center mass of these
two partons and the center mass of the remaining partons. Note that these
relative transverse vectors are Lorentz invariant quantities so that their values
are the same as their values in a rest hadron.  Finally,  ${\bf R}_a$ is the center
mass coordinate of hadron A. 
The momentum and coordinate variables can  then be
replaced by the new variables by the following equations:
\be &&{\bf a}_{1}=x_{a1}{\bf A}+\xi_{a1}{\bf p}_{21a}-{\bf a}_{21},\non \\
&&{\bf a}_{2}=x_{a2}{\bf A}+\xi_{a2}{\bf p}_{21a}+{\bf a}_{21},\non \\ &&{\bf
p}_a=x_a{\bf A}-{\bf p}_{21a},\non \\ &&{\bf r}_{a1}={\bf R}_{a}+x_{a} {\bf
r}_{{\bar a}a}-\xi_{a2}{\bf r}_{21a},\non \\ &&{\bf r}_{a2}={\bf R}_{a}+x_{a}
{\bf r}_{{\bar a}a}+\xi_{a1}{\bf r}_{21a},\non \\ &&{\bf r}_a={\bf
R}_a-(x_{a1}+x_{a2}){\bf r}_{{\bar a}a}.
\label{eq: aapr}
\en   The Jacobian of these replacements is 1. Similarly  we can
define  the  momentum variables ${\bf b}_{21}$, ${\bf p}_{21b}$, and ${\bf
B}$ and  their conjugate coordinate variables${\bf r}_{21b}$, ${\bf r}_{{\bar b}b}$, 
and ${\bf R}_b$  for hadron B.  To express Eq. (\ref{eq: dsdb}) 
in terms of coordinate variables we  Fourier transform the wave
functions from   the transverse  momentum space to the coordinate space by using, for example$\\$
 $\psi({\bf a}_{21},{\bf p}_{21a};A)=1/(2\pi)^2\int d{\bf r}_{21a} d{\bf r}_{{\bar
a}a}{\rm exp}(i{\bf a}_{21}
\cdot{\bf r}_{21a}+i{\bf p}_{21a}\cdot{\bf r}_{{\bar a}a}){\tilde \psi}({\bf
r}_{21a},{\bf r}_{{\bar a}a};A)$.$\\$
The integration volume $\int d^2{\bf p}_a \int d^2{\bf a}_1\int d^2{\bf a}_2$ in Eq. (\ref{eq: dsdb}) can be replaced by $\int d^2{\bf p}_{a21}\int d^2{\bf a}_{21}\int d^2{\bf A}$.   Since we are interested in the experiments with observables $Q_i^+$, $Q_i^-$ of a few TeV, we can therefore  neglect the ${\bf p}_{a21}$,   ${\bf a}_{21}$, ${\bf A}$  dependence in $|T|^2$. Integrating Eq. (\ref{eq: dsdb}) over these transverse momenta we get the differential cross section as:   
\be {d\sigma\over {d^4Q_1 d^4Q_2}}&=&{d\sigma\over {d^2{\bf Q}_Td^2\Delta{\bf
Q}dQ^+_1 dQ^-_1dQ^+_2  dQ^-_2}}\n &=&{1\over 16s^4(2\pi)^{8}}{1\over
x^2_{a1}x^2_{a2}}{1\over x^2_{b1}x^2_{b2}} 
\sum_{s_{a1}s_{a2}s'_{a1}s'_{a2}}\sum_{s_{b1}s_{b2}s'_{b1}s'_{b2}}|T|^2\n &\t&
{1\over 2}\sum_{S_A}\sum_{S_{a}}{1\over 2}\sum_{S_B}\sum_{S_{b}}\int [dV_a]\int
[dV_b]
\non\\ &\times& \int d^2{\bf r}_{{\bar a}a} \int d^2{\bf r}_{21a} \int d^2{\bf
r}_{{\bar b}b}  \int d^2 {\bf r}'_{{\bar a}a} \int d^2{\bf r}'_{21a} \non \\
&\t& {\tilde \psi}_{am}(x_{a1},s_{a1};x_{a2},s_{a2};{\bf r}_{{\bar a}a};{\bf
r}_{21a};V_a,S_a) 
{\tilde \psi}_{bn}(x_{b1},s_{b1};x_{b2},s_{b2};{\bf r}_{{\bar b}b};{\bf r}_{21a};V_b,S_b)\non
\\ &\t& {\tilde \psi}^*_{am}(x_{a1},s'_{a1};x_{a2},s'_{a2};{\bf r}'_{{\bar a}a};{\bf
r}'_{21a};V_a,S_a)
{\tilde \psi}^*_{bn}(x_{b1},s'_{b1};x_{b2},s'_{b2};{\bf r}'_{{\bar b}b};{\bf
r}'_{21a};V_b,S_b)\non \\ &\t& {\rm exp}(i{\bf Q}_T[{\bf r}_{{\bar a}a}-{\bf
r}'_{{\bar a}a}+{1\over 2}(\xi_{1a}-\xi_{2a})
 ({\bf r}_{21a}'-{\bf r}_{21a})]) {\rm exp}[i\Delta{\bf Q}({\bf r}'_{21a}-{\bf
r}_{21a})].
\label{eq: dssd}\en  

It is again  easy to see from the overall delta function that 
for double parton scattering process, both the momenta ${\bf Q}_T$ and $\Delta{\bf Q}$ are of the order of parton's transverse momenta which are of a few MeV.  We can therefore neglect the ${\bf Q}_T$ and $\Delta{\bf Q}$ dependence in  $|T|^2$ and we can integrate Eq. (\ref{eq: dssd}) over  ${\bf Q}_T$ and $\Delta {\bf Q}$. These integrations give two delta functions $\delta^2({\bf r}'_{21a}-{\bf
r}_{21a})$ and $\delta^2({\bf r}_{{\bar a}a}-{\bf
r}'_{{\bar a}a})$. We see from the Appendix that the spin flip process for $|T|^2_{\rm gp}$ is negligible therefore we can set $s_{ai}=s'_{ai}$, $s_{bi}=s'_{bi}$. These results enable us to express the differential cross section in terms of the square of wave functions:  
\be  {d\sigma\over {dQ^+_1 dQ^-_1 dQ^+_2dQ^-_2}}&=&{1\over
16s^4(2\pi)^{4}}{1\over x^2_{a1}x^2_{a2}} {1\over x^2_{b1}x^2_{b2}}
\sum_{s_{a1}s_{a2}s_{b1}s_{b2}} |T|^2 \int d^2{\bf r}_{21a}
\non\\ &\t&  {1\over 2}\sum_{S_A}\sum_{S_a}\int [dV_a] \int d^2{\bf r}_{{\bar
a}a} \non |{\tilde \psi}_{am}(x_{a1},s_{a1};x_{a2},s_{a2}; {\bf r}_{{\bar a}a};{\bf
r}_{21a};V_a,S_a)|^2\n &\t&{1\over 2}\sum_{S_B}\sum_{S_{b}}\int [dV_b] \int
d^2{\bf r}_{{\bar b}b} |{\tilde \psi}_{bn}(x_{b1},s_{b1};x_{b2},s_{b2};{\bf r}_{{\bar
b}b};{\bf r}_{21a};V_b,S_b)|^2.
\label{eq: dssb}
\en  We can define the following function  
\be  
 f_a(x_{a1},x_{a2},{\bf r}_{21a},s_{a1},s_{a2})&=& {1\over
4x_{a1}x_{a2}}{1\over (2\pi)^6}{1\over 2}
\sum_{S_A}\sum_{S_{a}}\int [dV_a]\int d{\bf r}_{{\bar a}a}\non
\\ &\t&|{\tilde \psi}_{am}(x_{a1},s_{a1};x_{a2},s_{a2};{\bf r}_{{\bar a}a};{\bf
r}_{21a};V_a,S_a)|^2.
\label{eq: famx}\en    By the normalization equation Eq. (\ref{eq: mssm}) we can show that $f_a$ the distribution function for two partons with momentum fractions $x_{a1}$ and
$x_{a2}$ in a relative transverse distance ${\bf r}_{21a}$. By using these distribution functions in Eq. (\ref{eq: dssb}) we can express the cross section as
\be {d\sigma\over {dQ^+_1 dQ^-_1 dQ^+_2dQ^-_2}}&=& {d\sigma\over {{1\over
2}d{Q}^2_1 dy_1 {1\over 2}d{ Q_2}^2 dy_2}}\n &=&{(2\pi)^8\over s^4}{1\over
x_{a1}x_{a2} } {1\over x_{b1}x_{b2} }
\sum_{s_{a1}s_{a2}s_{b1}s_{b2}} |T|^2\n &\t& \int d^2{\bf r}_{21a}
f_a(x_{a1},x_{a2},{\bf r}_{21a},s_{a1},s_{a2}) f_b(x_{b1},x_{b2}, {\bf
r}_{21a},s_{b1},s_{b2}).
\label{eq: dsdr}\en If we assume that partons are weakly correlated and are
evenly distributed in a hadron,
 we can use a flat distribution for the hadron wave  function. For example,  for hadron A we have:
\be  &&{1\over 4x_{a1}x_{a2}(2\pi)^6}{1\over 2}\sum_{S_A}\sum_{S_a}\int [dV_a] |{\tilde \psi}(x_{a1},s_{a1};x_{a2},s_{a2};{\bf r}_{a1}; {\bf
r}_{a2};V_a,S_a)|^2\n &&= f_a(x_{a1},s_{a1};x_{a2},s_{a2}){\Theta(({\bf
r}_{a1}-{\bf R}_a)^2-R^2)\Theta(({\bf r}_{a2}-{\bf R}_a)^2-R^2)\over (\pi
R^2)^2}.
\label{eq: xfax}
\en                     
By using  this in Eq. (\ref{eq:  dsdr}) and by carrying out the  $d^2{\bf r}_{21a}$ integration, we get 
\be {d\sigma\over {dQ^+_1 dQ^-_1 dQ^+_2dQ^-_2}}={d\sigma\over {{1\over 2}d{
Q}^2_1 dy_1  {1\over 2}d{ Q_2}^2 dy_2}} &=& {(2\pi)^8\over s^4}{1\over
x_{a1}x_{a2} } {1\over x_{b1}x_{b2} }
\sum_{s_{a1}s_{a2}s_{b1}s_{b2}} |T|_{\rm dp}^2 {\kappa \over \pi R^2}  \n
&\t&f_a(x_{a1},x_{a2},s_{a1},s_{a2}) f_b(x_{b1},x_{b2},s_{b1},s_{b2}),
\label{eq: dsxx}\en where $\kappa \approx 0.5$ is a geometrical factor
associated with the overlapping area of  the two colliding hadrons. For the cross section of the di-lepton production, we can use the result  in Appendix A
for $|T|_{\rm dp}^2$ and get
\be {d\sigma\over {dm_1 dy_1 dm_2dy_2}}|_{y_1=y_2=0} =\sum_{a_1,a_2}{2\kappa 
{\hat e}^2_{a_1}{\hat e}^2_{a_2}\over \pi R^2}f(x_{a1},x_{a2})f(x_{b1},x_{b2}) ({8\pi 
\alpha^2\over 9m_1s}) ({8\pi \alpha^2\over 9m_2s}),
 \label{eq: dmdy}\en where $m_i=\sqrt {Q_i^2}=sx_{ai}x_{bi}$ is the invariant mass square of the lepton pairs and ${\hat e_{a_i}}$ is the fractional charge of parton $a_i$. The summation  is to sum over the flavors of parton $a_1$ and  $a_2$.  The factor 2 in Eq. (\ref{eq: dmdy}) comes from when we include the Feynman diagram in Fig.~3 and its cross diagram.  Equation (\ref{eq: dmdy}) enables us to calculate the 
double parton cross section by using two universal double distribution  functions and two individual " hard scattering" single parton cross sections
\newpage
\noindent{\bf V.~Gluon splitting process}
\vskip 0.3cm  
   In this section we discuss the 4-final particle cross section from the gluon splitting process. We can use Eq. (\ref{eq: dsxa}) for this process if we take the
interacting partons $a_1$ and $b_1$ as gluon partons. The corresponding partonic
scattering amplitude square is shown in Fig.~4. There are two quark loop integrations in this Feynman diagram and one of them  is 
\be  L&=&\int d^2{\bf k}dk^+dk^-{Tr[\gamma^\beta k\!\!\!\slash \varepsilon\!\!\!\slash_{\lambda b}(b\!\!\slash-k\!\!\!\slash)\gamma^\alpha(a\!\!\slash+k\!\!\!\slash-Q\!\!\!\!\slash_2)\varepsilon\!\!\!\slash_{\lambda a} (Q\!\!\!\!\slash_2-k\!\!\!\slash)] \over 
(2k^+k^--{\bf k}^2+i\epsilon) [2(k^+-Q_2^+)(k^--Q_2^-)-({\bf k}-{\bf
Q}_2)^2+i\epsilon]}\nonumber\\ &\t&  {1 \over [2(k^++Q_1^+)(k^--Q_2^-)-({\bf
k}-{\bf Q}_2)^2+i\epsilon] [2k^+(k^--Q_1^--Q_2^-)-{\bf k}^2+i\epsilon]}.
\label{eq: lkqk}\en
From the overall  delta function, if we assume that  all  the transverse momenta of the partons are negligible,  we have   
\be
 a_1\approx(Q_1^++Q_2^+,0,{\bf 0}),\quad\quad b_1\approx(0,Q_1^-+Q_2^-,{\bf 0}).
\label{eq: aqqb}\en

We first carry out the integration over $k^-$.  When $0<k^+<Q_2^+$, only the second $k^-$-pole in the denominator is located on the lower half complex plane. We can close the
contour on the lower plane and pick up a pole at
\be
 k_2^-=Q_2^-+{({\bf k}-{\bf Q}_2)^2\over 2(k^+-Q_2^+)}.\label{eq: k2qk}
\en
  When $-Q_1^+<k^+<0$, 
only the third $k^-$-pole in the denominator is located on the upper half
complex plane. We can close the contour on the upper plane and pick up a pole
at
\be  k_3^-=Q_2^-+{({\bf k}-{\bf Q}_2)^2\over 2(k^++Q_1^+)}.
\label{eq: k3qk}\en 
A plausible prejudgement is that the main contribution to
the loop integration comes from when all the momenta in the loop are almost on
their mass shell, $i.e.$, when
\be  k^-=Q^-_2,\quad\quad k^+=0, \quad\quad{\bf k}^2=0,\quad ({\bf k}-{\bf
Q}_2)^2=0.
\label{eq: kqkk}\en With this in mind, we can set $k_2^-=k_3^-=Q_2^-$ and get 
\be  L=\int d^2{\bf k}\int_{-Q_1^+}^{Q_2^+} dk^+{2 \pi iTr[\gamma^\beta k\!\!\!\slash \varepsilon\!\!\!\slash_{\lambda
b}( b\!\!\slash-k\!\!\!\slash)\gamma^\alpha(a\!\!\slash+k\!\!\!\slash-Q\!\!\!\!\slash_2)\varepsilon\!\!\!\slash_{\lambda a} (Q\!\!\!\!\slash_2-k\!\!\!\slash)]|_{k^-=Q_2^-}
\over  (2k^+Q_2^--{\bf k}^2+i\epsilon)2(Q_1^++Q_2^+)({\bf k}-{\bf Q}_2)^2 
(-2k^+ Q_1^--{\bf k}^2+i\epsilon)}.
\label{eq: ldkq}\en For $Q_1^+$ and $Q_2^+$ much larger than $|{\bf k}|$ and  $|{\bf k}-{\bf Q}_2|$, we can approximate the integration range of  $k^+$
  in Eq. (\ref{eq: ldkq}) to the whole space then integrate $dk^+$  by closing the contour in the  upper half plane and pick up a pole at 
$k^+={\bf k}^2/(2Q_2^-)$. Eq. (\ref{eq: ldkq}) is then
\be  L=\int d^2{\bf k} {(2\pi i)^2 Tr[\gamma^\beta (Q_2^-\gamma^+-{\bf k}\!\!\!\slash)\varepsilon\!\!\!\slash_{\lambda
b} (Q_1^-\gamma^++{\bf k}\!\!\!\slash)\gamma^\alpha  (Q_1^+\gamma^--{\bf k}\!\!\!\slash+{\bf Q\!\!\!\!\slash_2})  \varepsilon\!\!\!\slash_{\lambda a}(Q_2^+\gamma^-+{\bf k}\!\!\!\slash- {\bf Q\!\!\!\!\slash_2}]
 \over 
4(Q_1^++Q_2^+)(Q_1^-+Q_2^-){\bf k}^2({\bf k}-{\bf Q}_2)^2}. 
\nonumber\\
\label{eq: ldki}\en
There are many terms contribute to the denominator in Eq. (\ref{eq: ldki}). One can show that the terms proportional to $Q_1^+Q_1^-Q_2^+Q_2^-$ are equal to zero. The leading contributions to the denominator in Eq. (\ref{eq: ldki}) are terms proportional to $Q_i^+Q_i^-{\bf k}^i({\bf k}-{\bf Q}_2)^j$ and these terms make the integration over $d^2{\bf k}$ of Eq. (\ref{eq: ldki})  logarithmically divergent. One needs to introduce a  cut-off to regulate this divergence. However, the  cut-off is lost because of our previous approximation. To mimic  the physical cut-off we will use  the  ($2-2\epsilon$)-dimensional  regularization in the   ${\rm   M S}$ -scheme. In
dimensional regularization, one introduces a  factor  $\mu^{2\epsilon}$  to fix the
dimensionality. In order to simplify the fromula, one can  express $\mu$ 
   in terms
of   $\mu^{ }_{\rm MS}$  by the relation  $\mu^2_{\rm MS}=4\pi\mu^2e^{-\gamma^{ }_E}$  ,  in which  $\gamma_E$   is the Euler  constant.  The result of the $d^{2-2\epsilon}{\bf k}$ integration has a term with  ${1/ \epsilon}$ pole. The    prescription is to drop this pole term, which effectively cuts off the intergration at $|{\bf k}|\approx\mu_{ \rm MS}$. Since the physical cut-off is at ${\bf k}\approx{\bf Q}_2$ , we set $\mu_{\rm MS}\approx|{\bf Q}_2|$. Preceding  with this prescription we can get   
\be   \int d^{2-\epsilon}{\bf k} { {\bf k}^i({\bf k}-{\bf Q}_2)^j\over 
 {\bf k}^2({\bf k}-{\bf Q}_2)^2}=\pi [({1\over \epsilon}-\ln{{\bf Q}_2^2 \over \mu^2_{\rm MS}}){\delta^{ij}\over 2}-{{\bf Q}_2^i{\bf Q}_2^j\over {\bf Q}^2}].
\label{eq: dkkk}\en

After some tedious calculations we get  the leading contribution to the scattering matrix square integrated over the 4-lepton final momentum space as
\be  |T|^2=2{16\alpha_s^2\alpha^4\pi^2\over 9(2 \pi)^6} {1\over Q_1^2}{1\over Q_2^2}{\Delta {\bf Q}^4 ({1\over 2}\ln^2{\Delta {\bf Q}^2\over \mu^2_{\rm MS}}-\ln{\Delta {\bf Q}^2\over \mu^2_{\rm MS}}+1) \over
(Q_1^++Q_2^+)^2 (Q_1^-+Q_2^-)^2}.\nonumber\\
\label{eq: tqqq}\en
We have inserted a color factor $C_F=T_F^2/N_g=1/32$  for Fig.~4 into Eq. (\ref{eq: tqqq}), in which $T_F=1/2$ comes from $\delta_{\alpha\beta}T_F=Tr(t_\alpha t_\beta)$ and $N_g=8$ is the number of gluon color configurations.
Inserting $|T|^2$ to Eq. (\ref{eq: dsxa}) and assuming that for unpolarized hadrons the gluon  distributions  function of helicity
 1 is equal to  that of helicity -1, we get the cross section for the gluon splitting process
as
\be {d\sigma\over {d^2 \Delta{\bf Q }dQ^+_1 dQ^-_1 dQ^+_2dQ^-_2}}&=&
{d\sigma\over {{1\over 2}\Delta{\bf Q }d{Q}^2_1 dy_1 {1\over 2}d{ Q_2}^2
dy_2}}\n &=&{2\alpha_s^2\alpha^4(2\pi)^6\over 9s^2}{1\over x_{a1 } } {1\over x_{b1} }{1\over Q_1^2}{1\over Q_2^2}{\Delta {\bf Q}^4 ({1\over 2}\ln^2{\Delta {\bf Q}^2\over \mu^2_{\rm MS}}-\ln{\Delta {\bf Q}^2\over \mu^2_{\rm MS}}+1) \over
(Q_1^++Q_2^+)^2 (Q_1^-+Q_2^-)^2}\n &\t&   f_{ga}(x_{a1}) f_{gb}(x_{b1}).
\label{eq: dsfg}\en 
\newpage
\noindent{\bf VI.~Discussions}
\vskip 0.3cm
    If we use the result in the Apendix for $|T|^2_{\rm sp}$ and integrate $d^2 \Delta{\bf Q}$ for Eq. (\ref{eq: dsxa}) and Eq. (\ref{eq: dsfg})  up to a value $\Delta{\bf Q}={\bf q}_{max}$ in which $0<<{\bf q}_{max}<<Q_1^+,Q_1^-,Q_2^+,Q_2^-$, we can see that the cross sections for single parton and gluon splitting processes behave like
\be 
{d\sigma\over { dm_1 dy_1 dm_2dy_2}}&=&{1\over 3}\sum_{a_1}{16{\hat e}^4_{a1} \alpha^4\over 9s^2(2\pi)^2 x_{a1}x_{b1} } f_a(x_{a1}) f_b(x_{b1}){\pi{\bf q}_{max}^2\over \sqrt{Q_1^2}\sqrt {Q_2^2}}\n
&\t&  {(Q_1^++Q_2^+)(Q_1^-+Q_2^-)(Q_1^+Q_1^-+2Q_!^+Q_2^-+Q_2^+Q_2^-)\over (Q_1^+Q_1^-Q_2^+Q_2^-)},\label{eq: dqqq}
\en
\be  
{d\sigma\over { dm_1 dy_1 dm_2dy_2}}&=&
{2\alpha_s^2\alpha^4(2\pi)^6\over 9s^2x_{a1 }x_{b1} } f_a(x_{a1}) f_b(x_{b1}) {\pi{\bf q}_{max}^6 \over \sqrt {Q_1^2} \sqrt {Q_2^2} }\n &\t&  { [{1\over 2}\ln^2 ({\bf q}_{max}^2/ \mu^2_{\rm MS})-\ln({\bf q}^2_{max}/\mu^2_{\rm MS})+1]\over (Q_1^++Q_2^+)^2 (Q_1^-+Q_2^-)^2}.
\label{eq: dsln}\en

Since in these processes ${\bf Q}_T\approx$ 0,  $\Delta{\bf Q}\approx{\bf Q}_1\approx-{\bf Q}_2$, if we collect all the 4-particle events with transverse momenta ${\bf Q}_1$ or ${\bf Q}_2$ less than  ${\bf q}_{max}$, we should see the cross section of single parton process increases linearly with ${\bf q}^2_{max}$ and the cross section of the gluon splitting process increases approximately  as ${\bf q}^6_{max}$. However, from Eq. (\ref{eq: dmdy}) we see that the cross section of double parton process keeps at a constant value when  ${\bf q}^2_{max}$ increases. Also if we use the HERA result for the quark distribution function $f(x)\propto x^{-1.4}$ \cite{hera} and use  Eq. (\ref{eq:  xxbp}) and Eq. (\ref{eq: xaiq}) for the values of $x$'s, when we select the events with $Q_1^+, Q_2^+,Q_1^-,Q_2^+\approx Q$ we can see that  the differential cross section for single parton process hehaves like $s^{0.4}Q^{-6.8}{\bf q}_{max}^2$ and that  for gluon splitting process behaves like $s^{0.4}Q^{-10.8}{\bf q}_{max}^6$. For double parton cross section, if we assume that partons are evenly distributed in a hadron with radius $R$ and assume that the double parton distribution function $f(x_1, x_2)$ can be factorized into muliplication of two single parton distribution functions $f(x_1)f(x_2)$, we see the differential cross section in Eq. (\ref{eq: dmdy})  beheaves like $s^{0.8}Q^{-7.6}R^{-2}$. For experiments of searching double parton events  we suggest one  do  experiments at large colliding hadron center mass energy $s$ and collect the  events with  small ${\bf q}^2_{max}$ then check their cross section behavior on $Q$. For experiments of searching gluon splitting process we suggest one collect the events up to a large ${\bf q}^2_{max}$ value  and select the events with small $Q$. 

The significance of the  result of the double parton cross sections in this paper are two-fold. We can use Eq. (\ref{eq: dmdy}) for double parton cross section and fit it with the single parton distributions obtained from the deep inelastic scattering or Drell-Yan process then  find the  hadronsize $R$ or we can use the double parton experiments data to fit the new double parton distribution functions in a hadron. In both ways they give us a better understanding of the hadron structure.
\vskip 2.5 cm
\centerline{\bf ACKNOWLEDGMENTS}
\vskip 0.7 cm
   Special thanks to my advisor Prof. Davison E. Soper. In memory of my father who passed away druing the time of my working on this paper.
\pagebreak
\vskip 0.8 cm
\newpage
\noindent{\bf APPENDIX}
\vskip 0.3cm 
   In this appendix we calculate $|T|^2$ for single and double parton processes where  $|T|^2$ is the partonic matrix square integrated over the 4-final particle phase spaces:
\be
|T|^2=(\prod_{i=1}^4\int {d^3l_i\over  (2\pi)^3 2l_i^0}) \delta^4 (l_1+l_2-Q_1) \delta^4 (l_3+l_4-Q_2).
 |m|^2\label{eq: dlll}
\en
 The following result is essential for calculating $|T|^2$: 
\be
\int {d^3l_1\over   2l_1^0}\int {d^3l_2\over  
2l_2^0}\delta^4 (l_1+l_2-Q_1) 
 Tr(\gamma_{\alpha} \not l_1 \gamma_{\alpha'} 
\not l_2)={2\pi \over 3}(-g_{\alpha\alpha'}+{Q_{1\alpha}Q_{1\alpha'}\over
Q_1^2})Q_1^2.\label{eq: dlll}
\en
With this result, the leading Feynman diagram to $|T|_{sp}^2$ shown i Fig.~5 can be express as:  
\be |T|^2&=& { 16{\hat e}^4_{a_1}\alpha^4 \over 9(2\pi)^{6}} (-g_{\alpha\alpha'}+{Q_{1\alpha}Q_{1\alpha'}\over Q_1^2})
(-g_{\beta\beta'}+{Q_{2\beta}Q_{2\beta'}\over Q_2^2})  \non\\
 &\times& {Tr[(Q\!\!\!\slash_1-a\!\!\!\slash)\gamma^ {\alpha}a\!\!\!\slash\gamma^{\alpha'}
(Q\!\!\!\slash_1-a\!\!\!\slash)\gamma^{\beta'}b\!\!\!\slash\gamma^{\beta}]\over Q_1^2Q_2^2(Q_1-a)^2(Q_1-a)^2}\non\\
&+&{\rm cross\quad diagram\quad contribution}.
\label{eq: ttcr}
\en 
If we neglect the transverse momemta for partons, we can set $a\approx(Q_1^++Q_2^+,0,{\bf 0})$ and $b\approx(0,Q_1^-+Q_2^-,{\bf 0})$. By using these in Eq. (\ref{eq: ttcr}) we get   
\be |T|_{\rm sp}^2&=& { 16{\hat e}^4_{a_1}\alpha^4 \over 9(2\pi)^{6}} {4(Q_1^++Q_2^+)(Q_1^-+Q_2^-)(Q_1^+Q_1^-+2Q_1^+Q_2^-+Q_2^+Q_2^-)\over Q_1^+Q_1^-Q_2^+Q_2^-Q_1^2Q_2^2}.
\label{eq: tspe}
\en 

For double parton interaction, since it is a four-parton process in the feynman diagram, it is possible that
the inteference terms from the spin flip processes as show in Fig.~5. may occur. The spin flip contribution as shown in Fig.~6 is:
\be |T_1|^2&=&  {e^2e_q^2\over (2\pi)^6}{1\over Q_1^2} {2\pi \over 3}(-g^{ }_{\alpha\alpha'}+{Q_{1\alpha}Q_{1\alpha'}\over
Q_1^2})\non\\ &\times&{\overline \nu}_\theta(Q_1^-,{\bf Q}_1-{\bf
a},-1/2)(-ie_q\gamma_{\theta\beta}^\alpha) u_\beta(Q_1^+,{\bf a},1/2)\non \\
&\times&{\overline u}_{\beta'}(Q_1^+,{\bf a}',1/2)(ie_q\gamma_{\beta'\theta'}^{\alpha'})\nu_{\theta'}(Q_1^-,{\bf Q}_1-{\bf
a}',1/2) .
\label{eq: tqqi}
\en
To calculate $T_1$ we use spinors in Ref. \cite{soper}  for quarks moving in the $+z$ direction, they are
and use spinors in Eq. (\ref{eq: pppp}) for quarks moving in the $-z$ direction. The calculation showes that if we neglect the transverse momemta for partons, i.e , if we set $a_1\approx(Q_1^+,0,{\bf 0})$, $b_1\approx(0,Q_1^-,{\bf 0})$,
$a_2\approx(Q_2^+,0,{\bf 0})$ and $b_2\approx(0,Q_2^-,{\bf 0})$ the spin flip contribution 
is zero. With this result, we can calculate $|T|^2$ for double parton process by using
\be |T_{dp}|^2&=&{e^4e_q^4\over (2\pi)^{12}}{1\over Q_1^2 Q_2^2} {2\pi\over 3}(-g_{\alpha\alpha'}+
{Q_{1\alpha}Q_{1\alpha'}\over Q_1^2})
{2\pi\over 3}(-g_{\beta\beta'}+{Q_{2\beta}Q_{2\beta'}\over Q_2^2})\non\\ &\times&  
Tr[b\!\!\!\slash_1\gamma^
{\alpha}a\!\!\!\slash_1\gamma^{\alpha'}]
Tr[b\!\!\!\slash_2\gamma^{\beta}a\!\!\!\slash_2\gamma^{\beta'}]\non\\ &+&{\rm cross\quad diagram \quad contribution}.
\label{eq: ttrc}
\en
We get \be |T_{dp}|^2&=&2{{\hat e}_{a1}^2{\hat e}_{a2}^2\alpha^4\over (2\pi)^{6}}{16^2\over 9}.
\label{eq: ttrc}
\en

\pagebreak

\centerline{\bf FIGURE CAPTIONS}
\vskip 1.0cm
\begin{description}

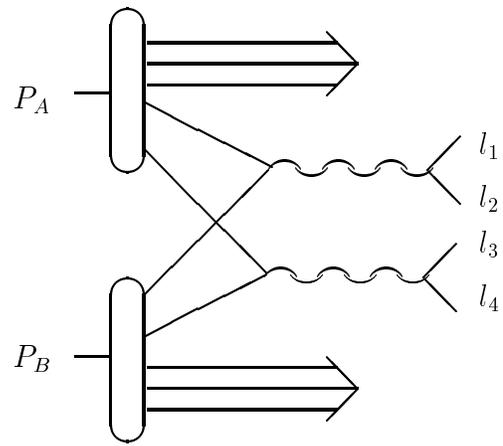
\item[Fig.~1] Double parton interaction of hadron-hadron scattering.

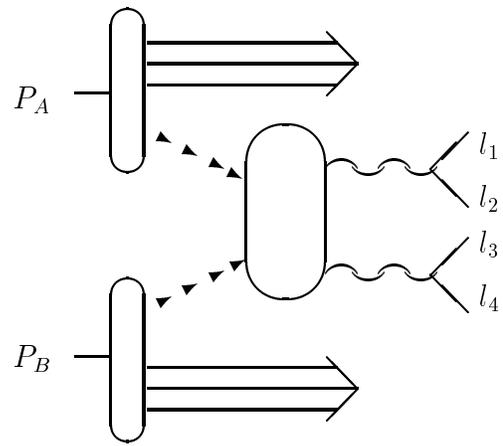
\item[Fig.~2] Single parton interaction of hadron-hadron scattering. The dot   
              line denotes quark or gluon partons.

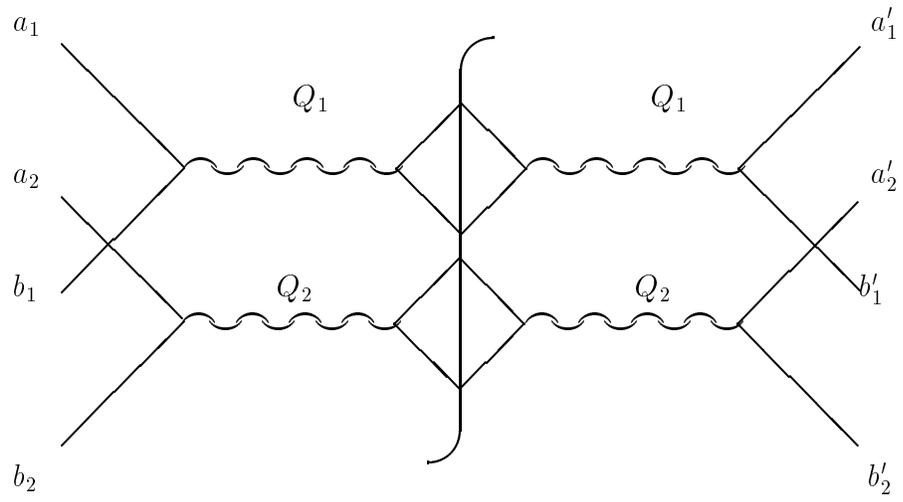
\item[Fig.~3] Partonic scattering matrix square of double parton interaction.

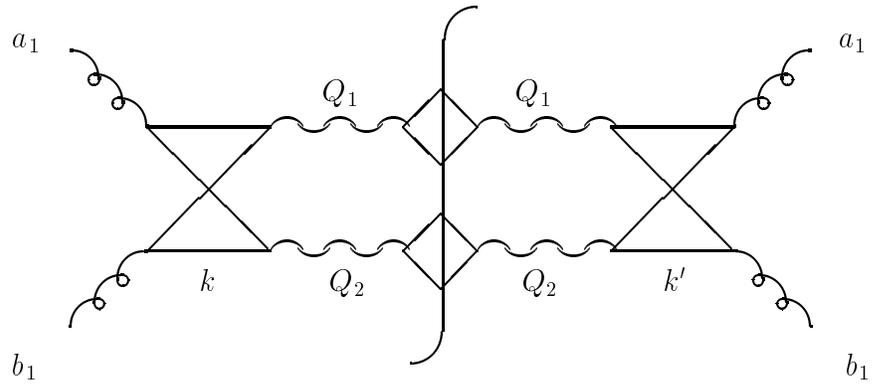
\item[Fig.~4] Partonic scattering matrix square of gluon splitting process.

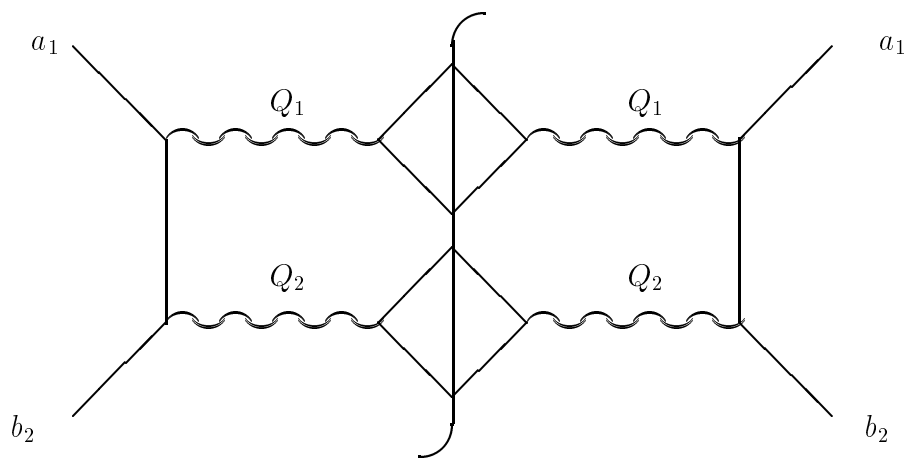
\item[Fig.~5] The leading Feynman diagram for single parton scattering.

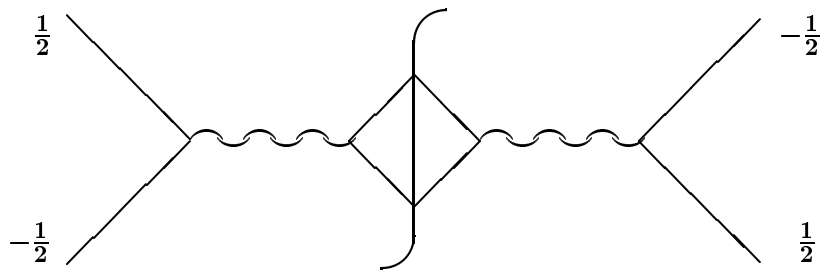
\item[Fig.~6] Spin-flip process contribution to double parton interaction.
  \end{description}
\label{eq: } 

\newpage
\centerline{\bf REFERENCES}
\vskip 0.3 cm
\begin{enumerate}
\bibitem{ua1} UA1 Collab., G. Arnison et al., Phys. Lett. B 158 (1985) 494; see
also W.  Scott, talk on Workshop on Physics in the 90's at the SPS Collider
(Zinal,Switzerland, June 1985CERN green report). UA2 Collab. J. A. Appel et al., CERN EP/85/189 (1985).


\bibitem{3j} J. Kuti and V. F. Weisskopf, Phys. Rev. D4 (1971) 3418; P. V. 
Landshoff and J. Polkinghorne, Phys. Rev.  D18 (1978)  3344; C. Goebel, D. M.
Scott and F. Halzen, Phys. Rev. D22 (1980) 2789; N. Paver, and D. Treleani,
Nuovo Cimento 70A  (1982) 215; 73 A  (1983) 392;  B. Humpert, Phys. Lett. 131B
(1983) 461; M. Jacob, CERN preprint TH-3515,3639(1983);  Z. Kunszt, Phys. Lett.
145B, (1984) 132;  N. Paver, and D. Treleani, Phys. Lett. 146B, (1984) 252; Z.
Phys. C28 (1985) 187; B. Humpert and R. Odorico,  Phys. Lett. 154B (1985) 211;
L. Ametller, N. Paver, and D. Treleani, Phys. Lett.  B169, (1986) 289. Z.Z. Kunszt ans W. J. Stirling, Phys. Lett. B171, 307 (1986).


\bibitem{4j} J. Gunion and Z. Kunszt, Phys. Lett. 159B, 167 (1985); S. Parke and 
T. Talor, Nucl. Phys. B269, 410 (1986); Z. Kunst, ibid. B271, 333 (1986); J.
Gunion and Z. Kunszt, Phys. Lett. 176B, 163; 477 (1986); J. Gunion and
J.Kalinowski, Phys. Rev. D 34, 2119 (1986); S. Parke and T. Talor, ibid. 35, 313
(1987);  F. A.  Berends and W. Giele, Nucl. Phys.  B294, 700 (1987); M. Mangano
and S. Parke  ibid. B299, 190 (1987). M. Mangano and S. Parke and Z. Xu, B 298,
673  (1988). ibid. B299, 190 (1987).


\bibitem{ua2} UA2 Collab. J. Alitti et al., Phys. Lett. B268, 145 (1991).

\bibitem{afs} AFS Collab., presented XXIII Inter. Conf. on High energy physics 
(Berkeley, July 1986);  ASF Collab. T. Akesson et al.,  Z. Phys. C34, 163(1987).
 
\bibitem{cdf} CDF Collab. F. Abe et al., Phys. Rev. D. 47, 4869 (1993).

\bibitem{soper} B. Kogut and Davison E. Soper, Phys. Rev. D1, 2901 (1970).

\bibitem{hera}  I. Abt. et al., Nucl. Phys. B407, 515 (1993).

\label{eq: } 
\end{enumerate}
\end{document}